\documentclass[trackchanges,twocolumn]{aastex701}

\begin{document}

\title{Fates of the sub-stellar objects (FOSSO) II. Evidence for Suppression of Metal Pollution in White Dwarfs by Close Substellar Companions}

\correspondingauthor{Di-Chang Chen; Bo Ma}
\email{chendch28@mail.sysu.edu.cn; mabo8@mail.sysu.edu.cn}

\author[orcid=0009-0009-0562-0202]{Zhangliang Chen}
\affiliation{School of Physics and Astronomy, Sun Yat-sen University, Zhuhai 519082, China}
\affiliation{CSST Science Center for the Guangdong-Hong Kong-Macau Great Bay Area, Sun Yat-sen University, Zhuhai 519082, China}
\email[]{chenzhliang@mail2.sysu.edu.cn}

\author[orcid=0009-0003-0759-6441]{Xin-Yue Zhang}
\affiliation{School of Astronomy and Space Science, Nanjing University, Nanjing 210023, China}
\affiliation{Key Laboratory of Modern Astronomy and Astrophysics, Ministry of Education, Nanjing 210023, China}
\email[]{xinyuez@smail.nju.edu.cn}

\author[orcid=0000-0003-0707-3213,sname='Chen']{Di-Chang Chen}
\affiliation{School of Physics and Astronomy, Sun Yat-sen University, Zhuhai 519082, China}
\affiliation{CSST Science Center for the Guangdong-Hong Kong-Macau Great Bay Area, Sun Yat-sen University, Zhuhai 519082, China}
\email{chendch28@mail.sysu.edu.cn}  

\author{Kejun Wang}
\affiliation{School of Physics and Astronomy, Sun Yat-sen University, Zhuhai 519082, China}
\affiliation{CSST Science Center for the Guangdong-Hong Kong-Macau Great Bay Area, Sun Yat-sen University, Zhuhai 519082, China}
\email{wangkj27@mail2.sysu.edu.cn}

\author[orcid=0000-0002-0378-2023]{Bo Ma}
\affiliation{School of Physics and Astronomy, Sun Yat-sen University, Zhuhai 519082, China}
\affiliation{CSST Science Center for the Guangdong-Hong Kong-Macau Great Bay Area, Sun Yat-sen University, Zhuhai 519082, China}
\email{mabo8@mail.sysu.edu.cn}

\author[orcid=0000-0002-6472-5348]{Ji-Wei Xie}
\affiliation{School of Astronomy and Space Science, Nanjing University, Nanjing 210023, China}
\affiliation{Key Laboratory of Modern Astronomy and Astrophysics, Ministry of Education, Nanjing 210023, China}
\email[]{jwxie@nju.edu.cn}

\author[orcid=0000-0003-1680-2940]{Ji-Lin Zhou}
\affiliation{School of Astronomy and Space Science, Nanjing University, Nanjing 210023, China}
\affiliation{Key Laboratory of Modern Astronomy and Astrophysics, Ministry of Education, Nanjing 210023, China}
\email[]{zhoujl@nju.edu.cn}

\begin{abstract}
Approximately 25–50\% of white dwarfs (WDs) exhibit metal absorption lines in their photospheres, interpreted as evidence of ongoing/recent accretion of planetary debris from remnant systems. 
Previous theoretical studies have suggested that massive, close-in substellar companion may prevent delivery of larger bodies via dynamical interactions, thereby reducing white-dwarf pollution.
However, no conclusive observational evidence has yet been established to confirm such a protective effect.
In this work, based on a sample of 17 white dwarf–substellar companion (1–75 $M_{\rm J}$) systems with reliable spectroscopic classifications, we find that white dwarfs hosting close substellar companions (orbital period $P < 5$ d) exhibit a metal-pollution fraction of $7.7^{+11.3}_{-4.0}\%$, which is suppressed by a factor of $5.75^{+3.24}_{-1.94}$ (corresponding to a protection efficiency of $87.2^{+3.4}_{-9.2}\%$) relative to single white dwarfs with a confidence level of 99.96\%.
In contrast, white dwarfs with wider companions show a metal-pollution fraction of approximately $25.0^{+24.0}_{-12.8}\%$, comparable to that of single white dwarf systems.
To interpret these results, we perform ensembles of N-body integrations and demonstrate that massive close-in substellar companions are capable of clearing 80\%–90\% of small-body contaminants.
The good consistency between the observational statistics and dynamical simulations provides strong evidence for suppressed metal pollution in white dwarfs with close companions, and offers insights into the long-term dynamical evolution of WD remanent systems. 
\end{abstract}



\section{Introduction}
\label{sec.intro}
White dwarfs (WDs) represent the final evolutionary stage of stars with masses less than about 8 $M_\odot$, which is the vast majority of stars in the Milky Way, including our Sun. 
Their atmospheres are expected to consist of nearly pure hydrogen and helium, as the extreme surface gravity causes heavier elements to sink out of the photosphere on timescales of days to millions of years, depending on the WD's effective temperature \citep{Koester09a}.
However, spectroscopic observations have revealed that approximately 25–50\% of all single WDs exhibit photospheric metal pollution \citep{Zuckerman03, Zuckerman10, Koester14}, with the total number of identified polluted WDs now exceeding 1,000 \citep{Coutu19}. 


The prevailing explanation for WD metal pollution is the recent or ongoing accretion of planetary debris \citep{Farihi09}. 
When main-sequence stars evolve off the giant branch, surviving planetary systems undergo dynamical reconfiguration due to stellar mass loss and tidal interactions \citep{Veras16}. 
Close-in planets/companions are likely engulfed as their host stars evolve into luminous red giants, while planets, companions and debris disks on wide orbits experience little interaction and could survive the post-main-sequence evolution \citep{chenzl26}.
Remaining planetesimals can be perturbed onto close-in orbits, where they are tidally disrupted to form debris disks that subsequently accrete onto the WD photosphere \citep{Jura03, Veras14, ChenDC19}. This process provides a unique window into the bulk composition of extrasolar planetesimals, complementing other exoplanetary characterization techniques \citep{Bonsor20}. 

While metal pollution is common among single WDs, the presence of close substellar companions—giant planets or brown dwarfs—may fundamentally alter the delivery of planetary material to the WD surface. 
Recent dynamical simulations demonstrate that close-in giant planets can dynamically eject or intercept highly eccentric asteroids and fragments, effectively shielding the WD from pollution \citep{ZhangXY26}. 
Observational evidence supporting this suppression mechanism remains limited, since only a handful of WDs known to host both confirmed close-in companions and measured atmospheric metal abundances \citep[e.g., ][]{Vanderburg20, Casewell18, Zorotovic22}.


In this work, we present a statistical analysis of metal pollution in WDs with close planetary and brown dwarf (BD) companions ($P<5$~day). 
By comparing the pollution incidence rates in these systems against WDs with wide or no companions, we provide observational evidence that close substellar companions significantly suppress metal pollution. 
Our results support the dynamical shielding scenario \citep{ZhangXY26} and provide constraints on the orbital separations at which this protection mechanism is effective. 
Specifically, in Section 2, we describe our sample selection and spectroscopic observations. 
Section 3 presents the  pollution rates of WDs with close and wide companions, deriving their protection effects of metal pollution comparing to WDs with no companions. 
In Section 4, we compare our results with the predictions of dynamical shielding scenario.
Finally, we summarize our main conclusions in Section 5.

\section{Sample selection}
\label{sec.sample}
Since the first discovery of a BD companion to the WD 0137-349 by \citet{Maxted06}, subsequent searches dedicating to WD-substellar systems have increased rapidly.
To date, approximately 30 substellar object candidates around WDs have been reported in the literature, of which 20 have been confirmed through multiple methods like RV measurements, spectral modeling or photometric eclipse detections.

\begin{figure}
    \centering
    \includegraphics[width=0.46\textwidth]{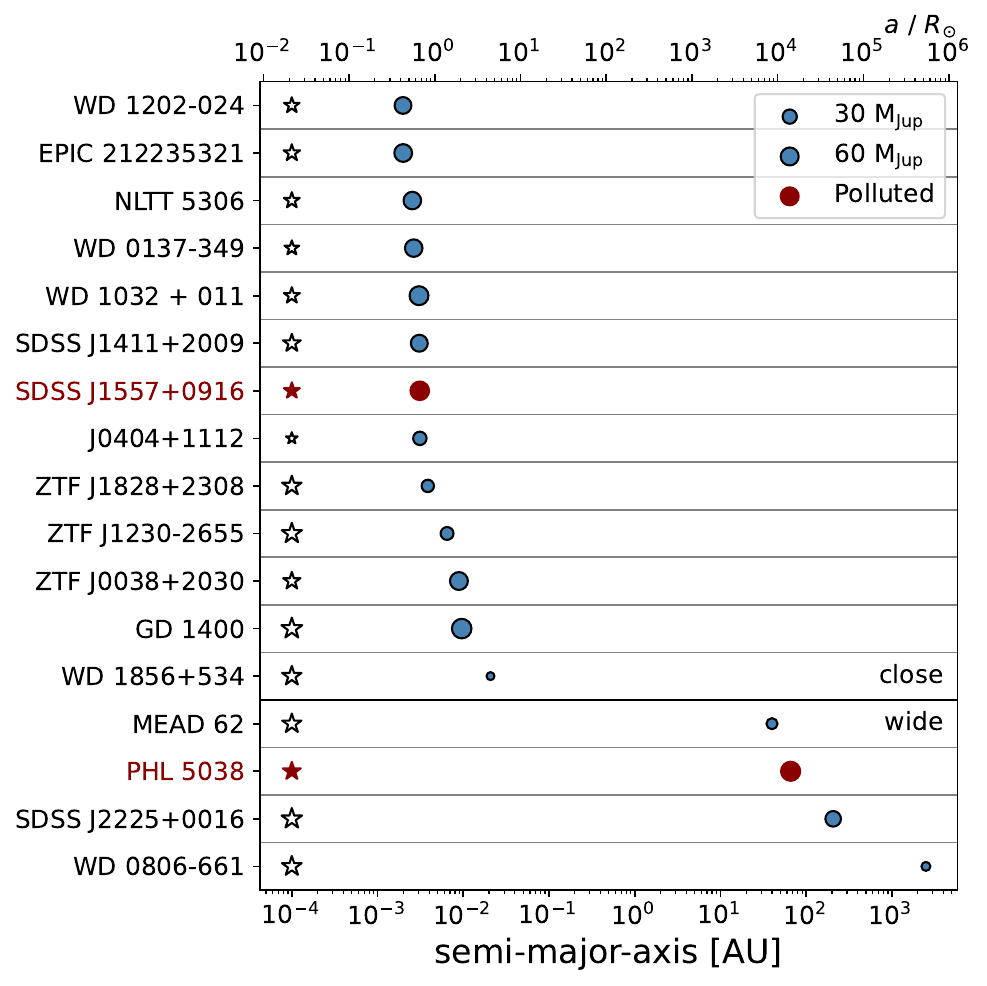}
    \caption{Well-constrained WDs with substellar companions (BDs or giant planets). The `close' and `wide' are defined as having an orbital period shorter or much longer than 5 days, respectively. Red markers denote the systems containing a polluted WD.}
    \label{fig:catalog}
\end{figure}

In this work, we focus on the systems with well-characterized orbital parameters that are suitable for assessing the WD metal pollution.
We compiled a sample of 17 WD-substellar companion systems from the literature and publicly available databases.
The following selection criteria have been applied:
\begin{itemize}
    \item The substellar companion mass should between 1--75~$\mathrm{M_{Jup}}$, corresponding to the mass region of giant planet and BD;
    \item {The WD must have a reliable spectroscopic classification with a spectral resolution of $R\gtrsim2000$, allowing its atmospheric type and significant metal pollution to be assessed.}
\end{itemize}
Systems were divided into two categories according to the orbital period: close systems (P $<$ 5 d), which are expected to have undergone common envelope evolution; and wide systems (P $\gg$ 5 d), where the two components are expected to have evolved independently.
We display our sample in Figure~\ref{fig:catalog}, where red markers denote the systems containing metal polluted WDs. {Their basic system parameters, WD spectroscopic instruments, and references are summarized in Table~\ref{tab:WDBD_catalog}.}

\begin{deluxetable*}{lcccccccccc}[!h]
\tablecaption{Catalog of known white dwarf--substellar companion systems} \label{tab:WDBD_catalog}
\tabletypesize{\scriptsize}
\tablewidth{0pt}
\tablehead{
\colhead{ID} & \colhead{WD mass} & \colhead{WD spT} & \colhead{WD Instr.} & \colhead{WD $T_{\mathrm{eff}}$} & \colhead{WD $t_{\mathrm{cooling}}$} & \colhead{SC mass} & \colhead{SC spT} & \colhead{$P_{\mathrm{orb}}$} & \colhead{Separation} & \colhead{Reference} \\
 & \colhead{[$M_{\odot}$]} & & &  \colhead{[K]} & \colhead{[Gyr]} & \colhead{[$M_{\mathrm{Jup}}$]} &  & \colhead{[day]} & \colhead{[AU]} &
} 
\startdata
EPIC 212235321 & 0.47 & DA & X-shooter & 24490 & 0.018 & 58 & L3 & 0.0474 & 0.002 & 1 \\
WD 1202-024 & 0.424 & DA & SDSS\tablenotemark{a} & 22640 & 0.05 & 49 & L0 & 0.0495 & 0.002 & 2, 3 \\
NLTT 5306 & 0.44 & DA & SDSS\tablenotemark{b}, X-shooter & 7756 & 0.71 & 56 & L4-L7 & 0.0707 & 0.0025 & 4, 5, 6 \\
WD 0137-349 & 0.39 & DA & UVES\tablenotemark{c} & 16500 & 0.25 & 55.5 & L8 & 0.0794 & 0.0026 & 7, 8 \\
SDSS J1411+2009 & 0.53 & DA & SDSS\tablenotemark{a} & 13000 & 0.27 & 50 & T5 & 0.0854 & 0.0031 & 9, 10, 11 \\
WD 1032 + 011 & 0.45 & DA & SDSS\tablenotemark{a} & 9904 & 0.455 & 70 & L5 & 0.0916 & 0.003 & 12 \\
SDSS J1557+0916 & 0.447 & DAZ & GMOS, X-shooter & 21800 & 0.033 & 66 & L3-L5 & 0.0947 & 0.0031 & 13 \\
ZTF J1828+2308 & 0.61 & DA & X-shooter & 15900 & 0.162 & 19.5 & $>$L4 & 0.112 & 0.0039 & 14 \\
J0404+1112 & 0.27 & DA & Keck/LRIS & 25000 & 0.0002 & 25 &  & 0.122 & 0.0031 & 15 \\
ZTF J1230-2655 & 0.65 & DA & X-shooter & 10000 & 0.726 & 22.1 & $>$L2 & 0.236 & 0.0065  & 14 \\
GD 1400 & 0.68 & DA & UVES\tablenotemark{c} & 11939 & 0.46 & 77.6 & L6–L7 & 0.416 & 0.0096 & 16, 17, 18 \\
ZTF J0038+2030 & 0.505 & DA & Keck/ESI & 10900 & 0.4 & 59.3 &  & 0.432 & 0.0089 & 19 \\
WD 1856+534 & 0.605 & DC & Literature\tablenotemark{d} & 4920 & 5.4 & 5.2 & Jupiter & 1.41 & 0.021 & 20, 21 \\
\tableline
MEAD 62 & 0.6 & DA & Literature\tablenotemark{e} & 5968 & 2.51 & 14 & Y & 1.19e+05 & 40 & 22 \\
PHL 5038 & 0.53 & DAZ & SDSS\tablenotemark{b}, X-shooter & 7525 & 1.14 & 72.4 & L8 & 2.69e+05 & 66 & 23, 24 \\
SDSS J2225+0016 & 0.66 & DA & SDSS\tablenotemark{b} & 10926 & 0.58 & 39 & L4 & 1.34e+06 & 207 & 25 \\
WD 0806-661 & 0.62 & DQ & IUE\tablenotemark{f} & 10250 & 0.67 & 7.3 & Y1 & 5.8e+07 & 2500 & 26, 27 \\
\enddata
\tablenotetext{}{\textbf{Reference.} 
(1).\cite{Casewell18EPIC};
(2).\cite{Rappaport17}; (3).\cite{Parsons17};
(4).\cite{Steele11}; (5).\cite{Steele13}; (6).\cite{Longstaff19};
(7).\cite{Maxted06}; (8).\cite{Burleigh06};
(9).\cite{Beuermann13}; (10).\cite{Littlefair14}; (11).\cite{Casewell18};
(12).\cite{Casewell20};
(13).\cite{Farihi17};
(14).\cite{Parsons25};
(15).\cite{deWit25};       
(16).\cite{Farihi04}; (17).\cite{Burleigh11}; (18).\cite{Casewell24};
(19).\cite{Roestel21};
(20).\cite{Vanderburg20}; (21).\cite{Limbach25};
(22).\cite{Albert25};
(23).\cite{Steele09}; (24).\cite{Casewell24};
(25).\cite{French23};
(26).\cite{Luhman11}; (27).\cite{Rodriguez11};
}
\tablenotetext{a}{SDSS spectroscopy from \citet{Kleinman13}.}
\tablenotetext{b}{SDSS spectroscopy from \citet{Eisenstein06}.}
\tablenotetext{c}{SPY (UVES) spectroscopy from \citet{Koester01} and \citet{Koester09b}.}
\tablenotetext{d}{Spectroscopy from multiple instruments in \citet{Vanderburg20}; no detectable metal lines, with stringent abundance upper limits.}
\tablenotetext{e}{Spectroscopy from multiple instruments in \citet{OBrien23}; MWDD indicates a spectral resolution of $R\sim4000$.}
\tablenotetext{f}{IUE ultraviolet spectroscopy from \citet{Subasavage09}.}
\end{deluxetable*}

\subsection{Close system}\label{sec.close}
For the close systems, we consider 9 confirmed WD-substellar systems and 3 WD--BD candidates.
Notably, 12 of the 13 systems host a hydrogen-atmosphere WD (DA-type).
The majority of these systems show no detectable photospheric metal absorption lines and no significant infrared excess attributable to circumstellar dust, suggesting the absence of ongoing accretion of planetary debris at currently observable levels. 

The only exception close WD--BD system is SDSS J155720.77+091624.6 (hereafter, SDSS~J1557+0916), which was first identified as a candidate debris disk system based on NIR photometry from \textit{UKIDSS} survey \citep{Steele11}.
Subsequent observations by \citet{Farihi17}, obtained with GMOS and X-shooter spectrographs, provided strong constraints on this system.
The data confirmed a close L5 dwarf companion and revealed photospheric metal absorption lines in the WD atmosphere, including Ca and Mg (Mg abundance [Mg/H] $\approx -4.5$, total mass accretion rate $\dot{M_1}\approx 6\times10^8~\mathrm{g~s^{-1}}$).
In addition, mid-infrared observations with \textit{Spitzer} also demonstrated a significant infrared excess that cannot be modeled by BD companion alone.
The excess emission was interpreted as thermal emission from circumstellar dust, which was inferred to reside in a circumbinary region exterior to the BD's orbit.
The features of SDSS~J1557+0916 indicate that it may have undergone a tidal disruption event involving a planetary body during its post-CE evolution.

\subsection{Wide system}\label{sec.wide}
{
Given the limited quality of the WD spectra in wide systems, we only consider 4 systems with reliable spectroscopic classifications, including 3 confirmed WD--BD binaries and 1 WD--BD candidate.
Owing to the small sample size, this sample is used for a qualitative comparison with the close systems and single WDs.
}

{Among these wide systems, one system (PHL~5038~AB) shows evidence of weak metal accretion or circumstellar disks.}
PHL 5038A is a $\mathrm{0.53~M_\odot}$ WD that shows a significant NIR flux excess compared to WD atmospheric model. 
Follow-up NIR observations by \citet{Steele09} confirmed that this excess arises from a BD companion at a projected separation of about 55 AU.
More recently, \citet{Casewell24} re-constrained the system parameters and investigated the atmospheric composition of the WD. 
Using new X-shooter spectroscopy, they measured a calcium abundance of [Ca/H] $\approx -9.4$ from the Ca II K line, and inferred a total mass accretion rate of $7.4\times10^6~\mathrm{g~s^{-1}}$, indicating the presence of photospheric metal pollution in the WD.

{We note that WD~0806-661 (GJ~3483) is a $\mathrm{0.62~M_\odot}$ DQ-type WD ($\mathrm{T_{eff}}\approx10250~\mathrm{K}$) with a co-moving, extremely faint Y-dwarf companion at a projected separation of $\sim2500$~AU, denoted as WD~0806-661~B \citep[GJ~3483~B, ][]{Luhman11, Rodriguez11, Voyer25}.
The WD exhibits carbon features in the ultraviolet spectrum with an abundance of [C/He] $\approx -5.5$ \citep{Subasavage09, Giammichele12}.
High-resolution UVES spectroscopy shows no evidence of additional metal absorption lines, indicating that the carbon features are unlikely to arise from external accretion, but instead originate from internal processes such as convective dredge-up of carbon from the core \citep{Pelletier86, Althaus09}.
}

\section{Observational analyse}
\label{sec.obs.results}

\subsection{Metal Pollution fraction for WDs systems}
In this subsection, we calculate the metal polluted fraction $F_{\rm MP}$ for WDs with close and wide companions.
Given $n_{\rm det}$ detections of metal pollution and the total number of systems $n_{\rm tot}$ studied, the posterior distribution of $F_{\rm MP}$ is given by:
\begin{equation}
    P (F_{\rm MP};a,b) = \frac{1}{B(a,b)} F_{\rm MP}^{a-1} (1-F_{\rm MP})^{b-1},
\end{equation}
where $B(a,b)$ is the beta function, with shape parameters $a=n_{\rm det}+1$ and $b=n_{\rm total}-n_{\rm det}+1$. 
The metal pollution fraction is reported as the maximum a posteriori (MAP) estimate from this Beta distribution, and the associated uncertainties are defined as the 68\% confidence interval centered on this value. 

\begin{figure}
    \centering
    \includegraphics[width=0.48\textwidth]{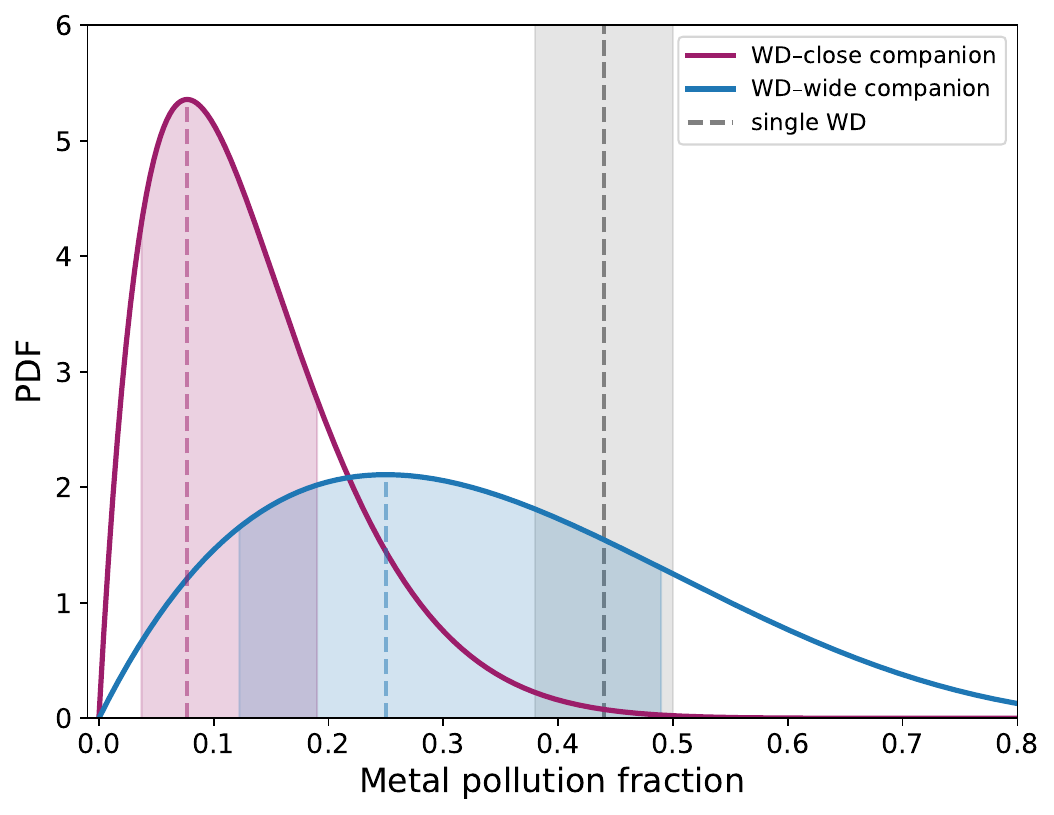}
    \caption{Probability distribution of metal-pollution fraction. The blue and purple lines represent the distribution for WDs with close and wide companions. The vertical gray region shows the metal-pollution fraction $44\pm6\%$ derived from a single WD sample \citep{2024ApJ...976..156O}.}
    \label{fig:Pollutionfraction}
\end{figure}

Figure~\ref{fig:Pollutionfraction} shows the probability distributions of the metal-pollution faction for WDs with close and wide companions.
As can be seen, WDs with close companions have a smaller metal-pollution fraction compared to those with wide companions.
Specifically, $F_{\rm MP}$ are $7.7^{+11.3}_{-4.0}\%$ and {$25.0^{+24.0}_{-12.8}\%$} for WDs with close and wide companions, respectively.

To assess whether the exist of substellar companions can affect the WD pollution, we compared with the metal-pollution fraction in single WD samples, which has been reported to be 25--50\% \citep[e.g.,][]{Zuckerman03, Zuckerman10}.
Here we adopt the metal-pollution fraction as $44 \pm 6\%$ (vertical gray band) derived from an independent HST/UV spectroscopic survey of single WDs \citep{2024ApJ...976..156O}.
To estimate the statistical significance, we assume that the metal pollution fractions follow the distributions described above, and generate 100,000 random realizations from the distributions for both the close companion and wide companion samples using Monte Carlo sampling method. We find that the metal-pollution fraction of WDs with close companions is lower than that of single WDs with a confidence level of 99.96\%.
In contrast, the pollution fraction for wide systems is a bit lower but statistically indistinguishable comparing to that of single WDs.

Previous studies have suggested that the accretion rates of WDs decrease with its cooling age \citep[e.g.,][]{Koester14,ChenDC19}.
However, this trend is unlikely to explain our results.  
The observable timescale of metal pollution can extend up to $\sim 8$~Gyr, whereas all systems in our sample have cooling ages below 7~Gyr, with the majority of the close-companion subsample being younger than 1~Gyr. 
Metal accretion, if present, should therefore be readily detectable for the white dwarfs in our sample, thus age-effects are unlikely to significantly affect the metal pollution fraction.
Moreover, even if the fraction of metal pollution were to decline with cooling age, the white dwarf–close companion systems in our sample are systematically younger than the white dwarf–wide companion systems, but they exhibit a lower observed pollution fraction, which is opposite to the expectation from a simple age-driven scenario. 
We therefore conclude that the reduced pollution fraction of white dwarfs with close companions is unlikely to be driven by age effects, and instead points toward a  suppression of debris accretion induced by close-in companions.

\subsection{Protection rate of close and wide companions}
To evaluate the influence of close and wide companions on the metal pollution, we then calculate the suppression factors and protection rates for the corresponding systems, which are defined as follows: 
\begin{equation}
    f_{\rm suppress} = \frac{n_{\rm single}}{n_{\rm det}},
\end{equation}
\begin{equation}    \label{eq: Fpro}
    F_{\rm protection} = 1-\frac{1}{f_{\rm suppress}},
\end{equation}
where $n_{\rm single} = n_{\rm total}\times F_{\rm MP}^{\rm single}$ is the predicted number of metal-polluted single WDs, and $n_{\rm det}$ is the observed number of metal-polluted WDs with a substellar companion.
Assuming a Poisson likelihood for the number of $n_{\rm det}$ detections and adopting a flat prior on the suppression factor $f_{\rm suppress}$, the posterior distribution of conditioned on follows a Gamma distribution as follows:
\begin{equation}
    P (f_{\rm suppress};c,d) = \frac{d^c f_{\rm suppress}^{c-1}*e^{-df_{\rm suppress}}}{\Gamma(c)},
\end{equation}
where $c=n_{\rm single}+1$ is the shape parameter and $d=1/n_{\rm det}$ is the rate parameter.
As mentioned before, the metal-pollution rate $F_{\rm MP}^{\rm single}$ is adopted as $44 \pm 6\%$.
Assuming a Gaussian distribution N(0.44, 0.06), we generated 1000 values of $F_{\rm MP}^{\rm single}$, corresponding to 1000 values of $n_{\rm single}$.
For each value, we combined $n_{\rm single}$ with $n_{\rm det}$ for WDs with close and wide companions, and performed Gamma distribution sampling with 1000 values per case.
This produced a total of 1,000,000 samples of the suppression factor $f_{\rm suppress}$ and the protection rate $F_{\rm protection}$.

\begin{figure}
    \centering
    \includegraphics[width=0.48\textwidth]{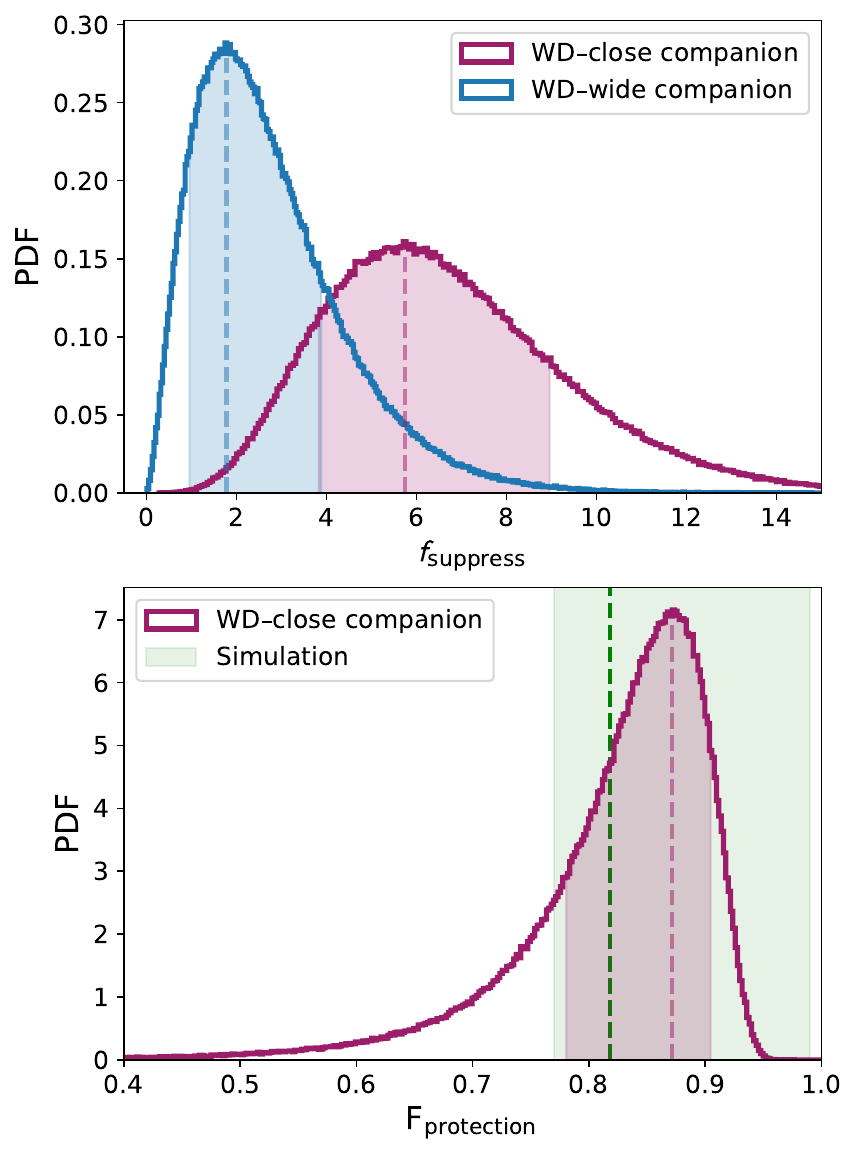} 
    \caption{Upper panel: Probability distribution of suppression factor for WDs with close (purple) and wide (blue) substellar companion. The shaded region indicates the 1-$\sigma$ confidence interval. Lower panel: Probability distribution of protection rate for WDs with close companion (purple). The green-shaded region represents the protection rate from N-body simulation with a system separation of 0.01~AU. The green dashed line shows the simulation result for the system SDSS~J1557+0916.}
    \label{fig: Protection rate}
\end{figure}

The probability density function (PDF) of $f_{\rm suppress}$ and $F_{\rm protection}$ are shown in Figure~\ref{fig: Protection rate}.
We find the maximum likelihood of suppression factor occurs at {$f^{\rm close}_{\rm suppress} = 5.75^{+3.24}_{-1.94}$ and $f^{\rm wide}_{\rm suppress} = 1.78^{+2.12}_{-0.84}$} for WDs with close and wide companions, respectively.
Since the suppression factor for wide systems is relatively small, we only show the protection rate for WDs with close companions $F^{\rm close}_{\rm protection}$, presented as PDF in the lower panel of Figure~\ref{fig: Protection rate}.
The distribution reaches the maximum likelihood at $F^{\rm close}_{\rm protection} = 87.2^{+3.4}_{-9.2}\%$.
We note that the peak protection rate is shifted higher than the value derived from the peak suppression factor using Equation~\ref{eq: Fpro}, because of the non-linear transfer.
Using the companion masses of the 13 close systems, we also calculate the range of protection rates from our simulations, shown as the green shaded region in Figure~\ref{fig: Protection rate}.
The green dashed line marks the protection rate corresponding to the companion in the only polluted close system SDSS~J1557+0916.
Details of simulation will be discussed in Section~\ref{sec.simulation}.

In addition, several WDs have been suggested to host close companions (e.g., SDSS J1604+1000, \citet{Irrgang21}; WD 0837+185, \citet{Casewell12}), but currently lack sufficient observational evidence. If these systems are included in the sample, the inferred suppression of metal-pollution by close companions would be even stronger.

\section{Theoretical simulation and comparison}
\label{sec.simulation}
To interpret the observed deficiency of metal pollution in WDs with close companions, we performed a series of numerical simulations following the methodology described in \citet{ZhangXY26}. Our goal is to quantify the protection rate provided by companions of various masses and orbital configurations against the infall of potential polluters, thereby providing a theoretical basis for comparison with our observations.

\begin{figure*}[t!]
    \centering
    \includegraphics[width=0.95\textwidth]{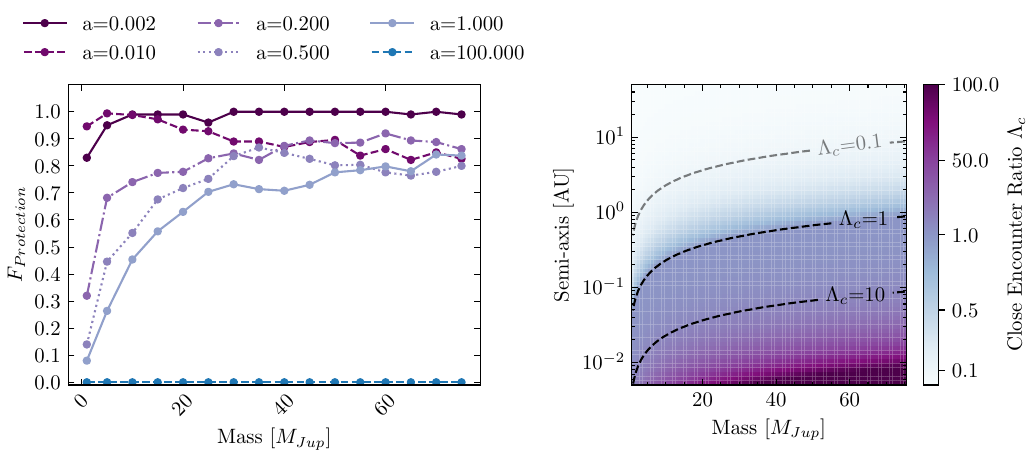}
    \caption{Simulation results for dynamical shielding scenario.
    Left panel: Protection rates ($F_{\rm protection}$) as a function of mass, with different line styles representing distinct semi-major axis values. 
    Right panel: Mapping the modified close encounter ratio in the planetary mass-semi major axis diagram.}
    \label{fig:bd_sim}
\end{figure*}

\subsection{Simulation setup}
We consider a representative system comprising a $0.6~M_{\odot}$ WD and a single companion. The companion mass ($M_c$) spans from $1$ to $75~M_{Jup}$, encompassing the regime from gas giant planets to brown dwarfs. To investigate the influence of orbital architecture, we sampled six representative semi-major axes ($a_c$) for the companion: 0.002, 0.01, 0.2, 0.5, 1 and 100~AU.

The potential sources of metal pollution are modelled as a population of test particles (representing asteroids or planetesimals) initially residing in an external reservoir, with their semi-major axes ranging from 1 to 10~AU. These particles are placed on highly eccentric orbits with periastron distances ($q$) that penetrate the WD's tidal radius ($R_{tidal}$), consistent with the standard delivery mechanism of circumstellar material into WD atmospheres. 
For a more comprehensive description of the numerical setup, we refer the reader to Section 2.1 of \citet{ZhangXY26}.

Utilizing the N-body integrator \texttt{REBOUND} \citep{rein_rebound_2012}, we evolved the system to track the dynamical evolution of the infalling test particles. 
A particle is classified as intercepted if its trajectory is significantly perturbed by the companion before reaching the WD's vicinity. 
We identify two primary channels for this protection mechanism:
\begin{enumerate}
    \item Direct Collision: The particle physically impacts the companion.
    \item Gravitational Scattering: The companion's gravitational influence either ejects the particle from the system or shifts its periastron distance such that $q > R_{\rm tidal}$, thereby preventing tidal disruption and subsequent accretion.
\end{enumerate}
The protection rate, denoted as $F_{protection}^{sim}$, is defined as the fraction of test particles that would have successfully polluted the WD in a single-star scenario but are diverted or removed by the presence of the companion.

\subsection{Simulation results}
We present the results of our numerical simulations and evaluate them against the theoretical framework, following a visualization approach consistent with Figure 6 of \citet{ZhangXY26}. The left panel of Figure \ref{fig:bd_sim} illustrates the protection efficiency as a function of the companion’s semi-major axis and mass. Complementing this, the right panel displays the close encounter ratio, $\Lambda_c$, derived from Equation (5) in \citet{ZhangXY26}, which serves as a theoretical proxy for the shielding effect. 
In agreement with the relations proposed by \citet{ZhangXY26}, our results confirm that more massive companions in tighter orbits provide more robust shielding for the WD. Specifically, while low-mass planets at 0.2~AU remain insufficient for effective shielding, a robust protection effect emerges for more massive BD companions ($M_c> 20~M_{Jup}$). 
For WD--BD separations within 0.5~AU, the simulated protection efficiency reaches approximately 80\%, which is in good quantitative agreement with the theoretical prediction of $\Lambda_c \approx 1$.

Furthermore, we observe that the protection efficiency exhibits a slight downward trend after reaching saturation at higher companion masses. {This subtle decline is linked to the companion's physical radius ($R_c$). Specifically, for higher companion masses, the empirical mass-radius relation used in our simulation \citep{muller2024} shows a declining $R_c$ with increasing mass, thereby weakening the efficacy of direct collisions between pollutants and the companion.} While the close encounter ratio, $\Lambda_c$, provides a robust first-order approximation {for gravitational scattering, this subtle downward trend, governed by the physical radius, falls outside its predictive scope.} However, while the adopted mass-radius relationship from \citet{muller2024} provides a reliable estimate for $R_c$ across the planetary to BD mass range, the actual radius of a BD is also sensitive to its age, metallicity, and internal composition \citep[e.g.,][]{Baraffe2003, Phillips2020}. Uncertainties in these stellar parameters can lead to a 10--20\% variation in the predicted BD radius. Such fluctuations propagate into the collision-driven protection fraction, introducing a corresponding uncertainty of approximately 5--8\% in the total protection efficiency for the innermost orbits ($a <$0.1~AU).

Our simulation specifically investigates the regime of extreme proximity, which is representative of our observational sample. For companions at $a \approx 0.01$~AU, the simulated protection efficiency reaches 80--100\%, which aligns well with the observed lack of pollution in such systems. We have plotted the simulated 0.01 au results against our observational data in Figure~\ref{fig: Protection rate}, showing a strong correlation between the presence of a close-in massive companion and the suppression of accretion.

Previous studies have proposed that massive, close-in planets can also block inward transport of small particles driven by radiative forces (e.g., Poynting-Robertson drag and the Yarkovsky effect), after the incoming asteroid has already been fully fragmented by tidal forces and mutual collisions \citep{veras2022orbit}.
However, in most systems, potential pollutants—such as rocky asteroids on rapidly infalling trajectories—are likely to be ejected by the massive, close-in companion long before these long-term radiative or fragmentation processes can take effect. 
Thus, the dynamical clearing by the companion remains the dominant mechanism, effectively intercepting the mass flux before a stable disk can be established.

Conversely, for the wide-companion group where the semi-major axis $a > $40~AU, our simulations show zero protection effect. Since the vast majority of rocky pollutants in our model originate from the inner planetary system ($a < $10~AU), these distant companions exert negligible dynamical scattering on the inward-bound material. Consequently, wide companions cannot prevent the WD pollution process, explaining why metal lines are frequently observed in WDs with distant companions.

\section{Summary}
\label{sec.summary}
Observations show that about 25--50\% of single WDs exhibit metal-pollution in their atmospheric spectrum, indicating ongoing or recent accretion of heavy elements.
These metals are generally attributed to debris produced by tidally disrupted event of planetary systems.
Recent studies have suggested that massive companions in close orbits around WDs may perturb the small-sized debris, and thereby reduce the fraction of WD pollution.

To test protective effect, we compiled a sample of well-constrained WD--substellar companion systems (Figure~\ref{fig:catalog}, Table \ref{tab:WDBD_catalog}) and found that the metal-pollution fraction of WDs with close companions is only $7.7^{+11.3}_{-4.0}\%$,  which is lower than that of single WDs with a 99.96\% confidence level (Figure~\ref{fig:Pollutionfraction}).
We then performed posterior-based Monte Carlo simulations and the derived suppression factor and protection rate for close companions are $5.75^{+3.24}_{-1.94}$ and $87.2^{+3.4}_{-9.2}\%$, respectively (Figure~\ref{fig: Protection rate}).
To interpret these results, following the method in \citet{ZhangXY26}, we then perform N-body simulations to quantify the protective influence of close-in planets on white-dwarf pollution by asteroids approaching on near parabolic orbits (Figure~\ref{fig:bd_sim}).
For close companions in our sample, dynamical shielding scenario yield a protection efficiency $\sim 90\%$, which is well consistent with the observation.
The above results present the first conclusive evidence that close companions to WDs can influence the long-term evolution of remnant planetary systems, and protect the WDs from metal pollution.

Nevertheless, the current sample of WD--substellar companion systems remains limited.
{In addition, other mechanisms may also contribute to the suppressed pollution fraction observed in close systems. Substellar objects formed within the protoplanetary can significantly perturb the disk structure by opening gaps and reducing the surface density \citep[e.g.,][]{Bryden99}. This may suppress the formation of other planets and planetesimals, thereby reducing the available reservoirs for subsequent WD pollution. Furthermore, non-isotropic mass loss during the common envelope phase may induce a natal kick and initiate rapid accretion \citep{Akiba24}, potentially depleting part of the reservoir before or shortly after the WD phase. Despite these possibilities, the low pollution fraction in close WD--substellar systems remains statistically significant in our current sample.}
Further observation results from surveys like Gaia \citep{GaiaMission16} and Earth 2.0 \citep{ETWhitePaper} are expected to significantly expand the sample size, allowing more verification of our conclusions and providing new insights into the long-term fate of planetary systems.

 
\begin{acknowledgments}
This work is supported by the National Key R\&D Program 
of China (Grant No. 2024YFA1611803) and the the National Natural Science 
Foundation of China (Grant No.12403071).
\end{acknowledgments}





\bibliography{sample701}{}
\bibliographystyle{aasjournalv7}

\end{document}